\documentclass[12pt,thmsa]{article}
\usepackage{amsmath}
\usepackage{graphicx}

\input{tcilatex}

\begin{document}

\begin{center}
{\Huge Topology and Turbulence}
\end{center}

\vspace{1pt}

\begin{center}
\textbf{R. M. Kiehn}

Mazan, France

rkiehn2352@aol.com

http://www.cartan.pair.com\vspace{1pt}\bigskip
\end{center}

\begin{quote}
\textbf{\vspace{1pt}Abstract: \ }Over a given regular domain of independent
variables \{x,y,z,t\}, every covariant vector field of flow can be
constructed in terms a differential 1-form of Action. The associated Cartan
topology permits the definition of four basic topological equivalence
classes of flows based on the Pfaff dimension of the 1-form of Action.
Potential flows or streamline processes are generated by an Action 1-form of
Pfaff dimension 1 and 2, respectively. Chaotic flows must be associated with
domains of Pfaff dimension 3 or more. \ Turbulent flows are associated with
domains of Pfaff dimension 4. \ It will be demonstrated that the
Navier-Stokes equations are related to Action 1-forms of Pfaff dimension 4.
\ The Cartan Topology is a disconnected topology if the Pfaff dimension is
greater than 2. \ This fact implies that the creation of turbulence (a state
of Pfaff dimension 4 and a disconnected Cartan topology) from a streamline
flow (a state of Pfaff dimension 2 and a connected topology) can take place
only by discontinuous processes which induce shocks and tangential
discontinuities. On the otherhand, the decay of turbulence can be described
by continuous, but irreversible, processes. Numerical procedures that force
continuity of slope and value cannot in principle describe the creation of
turbulence, but such techniques of forced continuity can be used to describe
the decay of turbulence.

\bigskip
\end{quote}

\section{INTRODUCTION}

The turbulence problem of hydrodynamics is complicated by the fact that
there does not exist a precise definition of the turbulent state that is
universally accepted. \ The visual complexity of the turbulent state leads
to the assumption that the phenomenon is in some way random and statistical.
It is however a matter of experience that a real viscous liquid which is
isolated from its surroundings, after being put into a turbulent state will
decay into a streamline state and ultimately a state of rest. This process
of decay is apparently continous

Although the turbulent state is intuitively recognizable, only a small
number of properties necessary for the turbulent state receive the support
of a majority of researchers:\medskip

\begin{enumerate}
\item  A turbulent flow is three dimensional

\item  A turbulent flow is time dependent.

\item  A turbulent flow is dissipative.

\item  A turbulent flow may be intermittent.

\item  A turbulent flow is irreversible.\medskip
\end{enumerate}

Still, a precise mathematical definition of the turbulent state has not been
established.

It sometimes is argued that a turbulent flow is random, a quality that has
been reinforced by successes of certain statistical theories in describing
average properties of the turbulent state. Recent advances in non-linear
dynamics indicate a sensitivity to initial conditions can lead to
deterministic ''chaos'', a quality which visually has some of the intuitive
features often associated with turbulence. Moreover, the theory of
non-linear dynamics has led to several new suggestions that describe the
route to turbulence.

However, there are fundamental differences between irreversibility, chaos,
and randomness that suggest that the turbulent state is not simply a chaotic
regime. Similar statements can be made about the transition to turbulence
and the formation of ''coherent structures'' in turbulent flows. Although
the eye easily perceives the antithesis to turbulence as being a steady
streamline flow, the transition from the pure streamline state to the
turbulent state is complicated by the fact that there may be intermediate
chaotic states in between an initial state of rest (or steady integrable
streamline flow) and a final state of turbulent flow. The objective of this
article is to focus attention on and apply topological methods, rather than
the customary geometric or statistical methods, to the problem of defining
the transition to turbulence and the turbulent state itself.

In hydrodynamics, it is generally accepted that the non-turbulent state
seems to be described adequately in terms of solutions to the Navier-Stokes
equations. The description of the turbulent state is not so clear, and to
this author the reason may be due to a lack of definition of what is the
turbulent state. As Ian Stewart [1988] states, ''the Leray Theory of
turbulence...asserts that when turbulence occurs, the Navier-Stokes
equations break down. ... turbulence is a fundamentally different problem
from smooth flow.'' In other words, one option would be that the turbulent
state is not among the solutions to the Navier-Stokes equations. Counter to
this option, and more in concurrence with the spirit of this article, would
be the inclusion of discontinuous solutions to the Navier-Stokes equations
into the class of evolutionary flows under consideration. \ However it is
possible to show that there exist continuous solutions to the Navier-Stokes
equations which are thermodynamically irreversible. \ Such solutions are
irreducibly four dimensional and can serve as candidates decscribing the
decay \ of turbulence. \ It is possible to demonstrate that the creation of
turbulence cannot be described by a continuous process.

The Kolmogorov theory of turbulence [Kolmogorov, 1941] is a statistical
option motivated by the assumption that the turbulent state consists of
''vortices'' of all ''scales'' with random intensities, but otherwise it is
a theory which is not based upon the Navier-Stokes equations explicitly. The
wavelet theory of Zimin [1991] is a method that does use a specific
decomposition of the solutions to the Navier-Stokes equations, and a
transformation to a set of collective variables, which mimic the Kolmogorov
motivation of vortices of all scales. However from a topological view,
scales can not have any intrinsic relevance.

The Hopf-Landau theory [Landau 1959] claims that the transition to
turbulence is a quasi-periodic phenomenon in which infinitely many periods
are sequentially generated in order to give the appearance of randomness. In
contrast, the Ruelle-Takens theory [Berge, 1984] describes the transition to
turbulence not in terms of an infinite cascade, but in terms of a few
transitions leading to a chaotic state defined by a strange attractor. In
other words, it would appear that the presence of a strange attractor
defines the turbulent state, but again there is a degree of vagueness in
that the ''strange attractor'' is ill-defined. A basic question arises: ''Is
chaos the same as turbulence?'' [Kiehn, 1990b].

In this article, topological methods will be used to formulate a definition
of the turbulent state. The method, being topological and not geometrical,
will involve concepts of scale independence, a concept in spirit recently
utilized by Frisch [1991] in an approach based on multi-fractals. However,
at the time of writing of the Frisch article, a specific connection between
fractals and solutions to the Navier-Stokes equations was unknown. Only
recently has the suggestion been made that those special characteristic sets
of points upon which the solutions to the Navier-Stokes equations can be
discontinuous, and which at the same time are minimal surfaces, may be the
generators of fractal sets [Kiehn, 1992].

A guiding feature of this article will be the idea that the abstract Cartan
topology of interest is refined by the constraints imposed by the
Navier-Stokes equations on the equations of topological evolution. Questions
of smoothness and continuity are always related to a specific choice of a
topology, and the topology chosen in this article is the topology generated
by a constrained Cartan exterior differential system [Bryant, 1991]. The
vector field solutions to the constrained topology (the Navier-Stokes
equations) fall into four equivalence classes that characterize certain
topological properties of the evolutionary system. Most of the mathematical
details are left to the Appendix, with the primary discussion presented in a
qualitative manner. The principle key feature of the constrained topology is
that two of the equivalence classes imply that the Cartan topology induced
by the vector flow field is disconnected, while the other two equivalence
classes imply that the induced topology is connected.

According to the theory presented herein, the key feature of the turbulent
state is that its representation as a vector field can only be associated
with a disconnected Cartan topology, while the globally streamline
integrable state is associated with a connected Cartan topology. It is a
mathematical fact that a transition from a connected topology to
disconnected topology can take place only by a discontinuous transformation.
This discontinuous transformation is typically realized as a cutting or
tearing operation of separation. However, a transition from a disconnected
topology to a connected topology can take place by a continuous, but not
reversible, transformation of pasting. The fundamental conclusion is that
the creation of the turbulent state is intrinsically different from the
decay of the turbulent state, but both processes involve topological change.

\section{THE CARTAN TOPOLOGY}

To come to grips with the topological issues of hydrodynamics it is
necessary to be able to define a useful topology in a way naturally suited
to the problem at hand. On the set \{x,y,z,t\} it is possible to define many
topologies. Indeed, for many evolutionary systems, particularly those which
are dissipative and irreversible, the topology of the initial state need not
be the same as the topology of the final state. Hence, not only is it
necessary to construct a topology for a hydrodynamic system at some time, t,
it is also necessary that the topology so constructed be a dynamical system
in itself. The transition to or from turbulence will involve topological
change.

As mentioned above, topological change can be induced either by a process
that is discontinuous, or by a process that is continuous, but not
reversible [Kiehn 1991a]. In this article attention is focused mainly on
those processes that are continuous but irreversible. Recall that continuity
is a topological property (not a geometrical property) that is defined in
terms of the limit points of the initial and final state topologies. The
methods used to define the topology used herein are based on those
techniques that E. Cartan developed for his studies of exterior differential
systems [Bishop, 1968]. For the hydrodynamic application it will be assumed
that the evolutionary physical system (the fluid) can be defined in terms of
a 1-form of Action built on a single covariant vector field. Typical of
Lagrangian field theory, the equations of evolution are determined from a
system of Pfaffian expressions constructed from the extremals to the 1-form
of Action when subjected to constraints. The resultant Pfaffian system
corresponds to a system of partial differential equations of evolution. The
extremal process is equivalent to the first variation in the theory of the
Calculus of Variations. The Action 1-form, A, may be composed from the
functions that make up the evolutionary flow field itself, as well as other
functions on the set \{x,y,z,t\}. A unreduced typical format for the Action
1-form on the set \{x,y,z,t\} would be given by the expression, 
\begin{equation}
A=A_{x}dx+A_{y}dy+A_{z}dz+A_{t}dt,
\end{equation}
where the functions$\;A_{k}$ are constructed from the components of the flow
field, and other functions. Cartans idea was to examine the irreducible
representations of this 1-form of Action. For example, in certain domains,
the action may be represented by the differentials of a single function, $%
A=d\phi $. In other cases, the representation might require two functions, $%
A=\alpha d\beta $, and so on. Given an arbitrary action, $A$, how do you
determine its irreducible represention? Is a given Action reducible? The
answer to the last question is given by the Pfaff dimension or class of the
1-form $A$. The Pfaff dimension is computed by one differential and several
algebraic processes producing a sequence of higher and higher order
differential forms. The construction of these forms is detailed in the next
section, but the remark to be made here is that these objects may be used to
construct a topological basis, and thereby, a topology.

In short, a point set equivalent to the Cartan topology can be defined in
terms of those functions that form the components of the covariant vector
field use to define the Action 1-form, the first partial derivatives of
these functions, and their algebraic intersections. The details of this
coarse Cartan topology over a space, \{x,y,z,t\}, are given in Appendix A,
with the refinement that constrains the coarse topology to yield those
evolutionary fields that are solutions to the Navier-Stokes equations.

\section{PFAFF DIMENSION}

The topological property of dimension is the key feature that distinguishes
precisely four Cartan equivalence classes of covariant vector fields on the
set \{x,y,z,t\}. The idea of Pfaff dimension (called the class of the
Pfaffian system in the older literature [Forsyth, 1953] ) is related to the
irreducible number of functions which are required to describe an arbitrary
Pfaffian form, in this case the 1-form of Action. An Action 1-form that can
be generated globally in terms of a single scalar field, and its
differentials, is of Pfaff dimension 1. Over space time, this single
function or parameter is often called the ''phase'' or ''potential''
function, $\phi (x,y,z,t)$. When the Action is of Pfaff dimension 1, then it
may be expressed in the reduced form, $A=d\phi .$

For the purposes described in this article, where the Action is defined in
terms of the flow field itself, the constraint of Pfaff dimension 1 implies
that the vector field representing the evolutionary flow is a ''potential''
flow. Such gradient vector flow fields represent a submersion from four
dimensional space-time to a parameter space of one dimension, the potential
function itself.

Actions constructed for flows that admit vorticity (but are completely
integrable in the sense that through every point there exists a unique
parameter function whose gradient determines the line of action of the flow)
can be represented by a submersive map to a parameter space of two
dimensions. The reduced format for the 1-form of Action is $A=\psi
(x,y,z,t)d\phi (x,y,z,t)$. Such integrable flows are defined as globally
laminar flows (in the sense that there exists a globally sychronizable set
of unique initial conditions, or parameters). Such flows are to be
distinguished from flows that may have, for example, re-entrant domains, and
are locally layered, but for which it is impossible to define a global 
\textit{connected} set of initial conditions. Globally laminar flows and
potential flows are of Pfaff dimension 2 or less, and are associated with a
Cartan point set topology which is \textit{connected}. This concept is to be
interpreted as implying that there exists an N-1 dimensional set which
intersects the flow lines in a unique set of points. (As Arnold says, the
field is of co-dimension 1.) For three dimensional space, N=3, this set is a
surface. For four-dimensional space, N=4, this set is a volume.

In a domain where the Cartan point set topology is connected (the 1-form of
Action is of Pfaff dimension 2 or less) it is possible to define a single
connected parameter of evolution which plays the role of ''phase''. In three
dimensions, the parametric value is called ''time''. This idea of a uniquely
defined global parameter N-1 surface is the heart of the Caratheodory theory
of equilibrium thermodynamics. For such systems, there are infinitely close
neighboring points which are not reachable by closed equilibrium processes.
The equilibrium process is defined to be a process whose trajectory is
confined to the N-1 integrable hypersurface. The two irreducible functions
of the integrable Pfaff representation are called temperature, T, and
entropy, S, in the Caratheodory setting. The parameter space is connected,
but a constant value of the parameter space does not intersect all points of
the domain. The evolutionary vector fields associated with completely
integrable Pfaffian systems are never chaotic [Schuster, 1984].

For a 1-form of Action that is associated with a disconnected point set
topology, such a globally unique parameterization as described above is
impossible. Those 1-forms of Action, if not of Pfaff dimension 2 (or less)
globally, do not satisfy the Frobenius complete integrability conditions
[Flanders,1963] The evolutionary vector fields associated with
non-integrable 1-forms can be chaotic. If the 1-form of Action is of Pfaff
dimension 3, then it has an irreducible representation as $A=d\beta +\psi
d\phi $. The three independent functions form a non-zero three form otf
topological torsion, $H=A\symbol{94}dA=d\beta \symbol{94}d\psi \symbol{94}%
d\phi $, and represents a covariant current of rank 3, with a dual
representation as a contravariant tensor density (the torsion current). The
torsion current has zero divergence on the domain of space \{x,y,z,t\} which
is of Pfaff dimension 3 relative to the Action A. This result implies that
the ''lines'' so generated by the solenoidal torsion current in \{x,y,z,t\}
can never stop or start within the domain interior. The lines representing
the topological torsion 4-current either close on themselves, or start and
stop on points of the boundary of the domain. The torsion lines never stop
in the interior of the domain where the Pfaff dimension is 3. Such a torsion
current does not exist in domains of Pfaff dimension 2 or less. Explicit
formulas for the torsion current will be given below.

For a four dimensional domain, the 1-form of Action may be of Pfaff
dimension 4, and the irreducible representation of the Action is given by
the expression, $A=\alpha d\beta +\psi d\phi $. Each of these functions is
independent, so the topological torsion current is of the form,

\begin{equation}
A\symbol{94}dA=\alpha d\beta \symbol{94}d\psi \symbol{94}d\phi +\psi d\phi 
\symbol{94}d\alpha \symbol{94}d\beta
\end{equation}
The topological torsion current is a 4 component vector field. However, the
divergence of this vector field is not necessarily zero! The lines of the
torsion current can start or stop in the \textit{interior} of the domain
when the Pfaff dimension is 4, but not when the Pfaff dimension is 3.

In four dimensions, a solenoidal vector field, if homogeneous of degree 0,
forms a minimal surface in space time. In fact, if a four dimensional vector
field can be represented by a complex holomorphic curve, then the field is
not only solenoidal, but also harmonic, and is always associated with a
minimal surface. It will be shown below, that in the hydrodynamics governed
by the Navier-Stokes equations, harmonic vector field solutions are not
dissipative, no matter what the value of the viscosity coefficient. For
dissipative irreversible systems, attention is therefore focused on systems
of Pfaff dimension 4, for which the torsion current is not solenoidal.

The theory presented in this article insists that irreversible turbulence
must be time dependent and irreducibly three dimensional. The idea of ''two
dimensional'' turbulence, for time dependent continuous flows, is
inconsistent, for such flows have a maximum Pfaff dimension of 3. Flows of
Pfaff dimension 3 can be chaotic, but they are deterministically reversible,
hence not turbulent. In agreement with the arguments expressed by Kida
[1989], the turbulent state is more than just chaos. A turbulent domain must
be of Pfaff dimension 4, for in space-time domains of Pfaff dimension 3, it
is always possible to construct flow lines that never intersect. Hence such
flow lines are always re-traceable, without ambiguity, and such flows are
not irreversible. In order to break time-reversal symmetry, and hence to be
irreversible, the flow lines must intersect in space-time such that they
cannot be retraced without ambiguity. Such a result requires that the Euler
characteristic of the four dimensional domain must be non-zero, for then it
is impossible to construct a vector field without intersections. The Euler
characteristic of space-time is only non-zero on domains of Pfaff dimension
4. Hence, the Pfaff dimension of turbulent domains on \{x,y,z,t\} must be 4,
while the chaotic domains need be only of Pfaff dimension 3 [Kiehn, 1991b].

\section{TOPOLOGICAL CONNECTEDNESS VS. GEOMETRIC SCALES}

In early studies of the turbulent state, from both the statistical point of
view and the point of view of the Navier Stokes equations, the geometric
concepts of large and small spatial scales, or short or long temporal
scales, have permeated the discussions. From a topological point of view,
length scales and time scales have no meaning. If things are too small, a
topologist stretches them out, and conversely. If turbulence is a
topological concept, then the ideas should be independent from scales. It is
interesting to note that the original Kolmogorov statistical analysis of the
turbulent state is now interpreted in terms of the multi-fractal concept of
scale invariance [Frisch, 1991].

A key feature of the disconnected Cartan topology is that the domain
supports non-null Topological Torsion, and is of Pfaff dimension 3 for
chaotic flows, and of Pfaff dimension 4 for turbulent flows. Suppose the
initial state is a turbulent state in which there exist disconnected
striated or tubular domains that are of Pfaff dimension 4 and are embedded
in domains of Pfaff dimension 2 or less. Then if the hydrodynamic system is
left to decay, these striated domains can decay by continuous collapse into
striations or filaments of measure zero. The Topological Torsion of the
striated domains cannot be zero. The size of the striated domains is not of
issue, but the existence of such domains with non-zero measure is of
interest, for if these domains do not exist, the flow is not chaotic and not
turbulent.

The geometric idea of small domains versus large domains of space and/or
time is transformed to a topological idea of connected domains versus
disconnected domains. Points in disconnected components are not reachable
[Hermann, 1968] in the sense of Caratheodory, hence are separated by ''large
scales'', while points in the same component are reachable, and hence are
separated by ''small scales'', compared to points in disconnected
components. It is the view of this article that the geometric concept of
scales is not germane to the problem of turbulence, but instead the basic
issue is one of connectedness or disconnectedness.

\section{TOPOLOGICAL TORSION}

The difference between chaotic flows and turbulent flows is that chaotic
flows preserve time reversal symmetry and turbulent flows do not. Chaotic
flows can be reversible, while turbulent flows are not. Both chaotic and
turbulent flows support a non-zero value of Topological Torsion tensor. As
constructed in the Appendix for the Navier-Stokes system, the Topological
Torsion 3-form on space-time, H, has 4 components, $\{\mathbf{T},h\}$ that
transform as the components of a third rank completely anti-symmetric
covariant tensor field, $H_{ijk}$. If the vector field used to construct the
Topological Torsion tensor is completely integrable in the sense of
Frobenius, then all components, $\{\mathbf{T},h\}$, vanish, and the Pfaff
dimension of the domain is 2 or less. For the Navier-Stokes fluid, the
torsion current is given by the engineering expression given in the appendix
as equation (20). As a third rank tensor field, the Topological Torsion
tensor is intrinsically covariant with respect to all coordinate
transformations, including the Galilean translation.

For engineers, the closest analog to the Topological Torsion tensor is the
charge-current, 4-vector density, $\mathbf{J}$, of electromagnetism. The
fundamental difference is that where the electromagnetic 4-current always
satisfies the conservation law, $div\;\mathbf{j}+\partial \rho /\partial t=0$%
, the Topological Torsion 4-current does not, unless the vector field use to
construct the Topological Torsion tensor is an element of an equivalence
class with Pfaff dimension less than 4. In other words, for the turbulent
state, the Topological Torsion tensor does not satisfy a local conservation
law, where for the chaotic state it does:$\medskip $

\begin{itemize}
\item  $div\;\mathbf{T}+\partial h/\partial t=0$ ... Pfaff Dimension 3,

(a necessary condition for chaos),

\item  $div\;\mathbf{T}+\partial h/\partial t\neq 0$ ... Pfaff Dimension 4,

(a necessary condition for turbulence).\medskip
\end{itemize}

The lines of torsion current, given by solutions to the system of first
order differential equations,

\begin{equation}
dx/\mathbf{T}_{x}=dy/\mathbf{T}_{y}=dz/\mathbf{T}_{z},
\end{equation}
can start or stop internally if the Pfaff dimension is 4, but only on
boundary points or limit points of the domain, if the Pfaff dimension is 3 .
If the vector field is of Pfaff dimension less than 4, then the integral
over a boundary of the Topological Torsion tensor is an evolutionary
invariant, but some care must be taken to insure that the integration domain
is a boundary, and not just a closed cycle. This result, which corresponds
to a global helicity conservation theorem, is independent of any statement
about viscosity. However, the invariance of the Topological Torsion integral
over a boundary for a Navier-Stokes fluid implies that integral of the
4-form of Topological Parity must vanish, which in turn implies that the
Euler index of the Cartan topology for this situation is zero. The
compliment of this idea leads to a variable Topological Torsion integral,
and the requirement that the Euler index is not zero for the irreversible
decay of the turbulent state. For if the Euler index is not zero, then every
vector field has at least one flow line with a singular point. If an
evolutionary parameter carries the process through the singular point, a
reversal of the process parameter will retrace the path only back to the
singular point uniquely. Subsequently, the return path then becomes
ambiguous, and the evolution is not reversible. Recall that a necessary
condition for a vector field to exist on a manifold without self
intersection singularities is that the Euler index of the manifold must
vanish. Hence a necessary condition for turbulence is that the Cartan
topology must be of Pfaff dimension 4, and the topological torsion is not
solenoidal.

It is remarkable to this author that experimentalists and theorists
(including the present author) have been so brain-washed by the dogma of
unique predictability in the physical sciences that they have completely
ignored the measurement and implications of the Topological Torsion tensor.
Although the solutions to a Pfaffian system of equations  is a problem that
has found use in the older literature of differential equations (where it is
known as the ''subsidiary'' system) [Forsyth 1959 p.95], its utilization in
applied dynamical systems, especially hydrodynamics, is extremely limited.
Of course, for uniquely integrable systems, the equations of topological
torsion are evanescent, and not useful. The very existence of the
Topological Torsion tensor is an indicator of when unique predictability is
impossible [Kiehn, 1976], and attention should be paid to the Pfaff
dimension of a physical system described by (1).

\section{TORSION WAVES}

One of the predictions of the Cartan topological approach is the fact that
for systems of Pfaff dimension 4 it is possible to excite torsion waves.
Torsion waves are essentially transverse waves but with enough longitudinal
component to give them a helical or spiral signature. In the electromagnetic
case, where such waves have been measured, they are represented by four
component quaternionic solutions to Maxwell's equations [Schultz, 1979;
Kiehn, 1991]. Such electromagnetic waves represent different states of left
or right polarization (parity) traveling in opposite directions. The wave
speeds in different directions can be distinct. In fluids, transverse
torsion waves can be made visible by first constructing a Falaco soliton
state [Kiehn 1991c] and then dropping dye near the rotating surface defect.
The dye drop will execute transverse polarized helical motions about a
guiding filamentary vortex that connects the pair of contra-rotating surface
defects. There is some evidence that torsion waves can appear as traveling
waves on Rayleigh cells [Croquette, 1989].

\section{THE PRODUCTION VS. THE DECAY OF TURBULENCE}

The Cartan topological theory predicts that the transition to turbulence
from a globally laminar state involves a transition from a connected Cartan
topology to a disconnected Cartan topology. From this fact it may be proved
that such transitions can NOT be continuous, but they may be reversible!
However, the theory also predicts that the decay of turbulence can be
described by a continuous transformation, but the transformation can NOT be
reversible.

The Cartan topology when combined with the Lie derivative may be used to
define partial differential equations of evolution [Kiehn, 1990], which
include the Navier-Stokes equations as a subset of a more refined topology.
However, if the Cartan topology is constructed from p-forms and vector
fields that are restricted to be C2 differentiable, then it may be shown
that all such solutions to the Navier-Stokes equations are continuous
relative to the Cartan topology. The \textit{creation} of the turbulent
state must involve discontinuous solutions to the Navier-Stokes equations,
which are generated only by shocks or tangential discontinuities, and
therefore are not describable by C2 fields. On the other hand, the \textit{%
decay} of turbulence can be described by C2 differentiable, hence
continuous, solutions which are not homeomorphisms, and are therefore not
reversible. In this article, the decay of turbulence by C0 and C1 functions
is left open.

Domains of finite Topological Torsion are topologically disconnected from
domains that have zero Topological Torsion. The anomaly that permits the
local creation or destruction of Topological Torsion is exactly the 4-form
of Topological Parity (see Appendix). If the Topological Parity is zero,
then the Topological Torsion obeys a pointwise conservation law. For a
barotropic Navier-Stokes fluid, the anomaly, or source term for the Torsion
current can be evaluated explicitly, and appears as the right hand side in
the following equation:

\begin{equation}
div\;\mathbf{T}+\partial h/\partial t=-2\nu \;curl\;\mathbf{v}\bullet
curlcurl\;\mathbf{v}.
\end{equation}

It is remarkable that for flows of any viscosity, the Topological Torsion
tensor satisfies a pointwise conversation law, and the integral over a
bounded domain is a flow invariant, if the vorticity vector field satisfies
the Frobenious integrability conditions, $curl\;\mathbf{v}\bullet curlcurl\;%
\mathbf{v}=0$. It would seem natural that the decay of turbulence would be
attracted to such interesting limiting configurations of topological
coherence in a viscous fluid. These limit sets can be related to minimal
surfaces of tangential discontinuities which can act as fractal boundaries
of chaotic domains [Kiehn 1992b, 1992c, 1993].

\section{SUMMARY}

The topological perspective of Cartan indicates that:

\begin{enumerate}
\item  A necessary condition for the turbulent state is that the flow field
must generate a domain of support which is of Pfaff dimension 4, and is to
be distinguished from the chaotic state, which is necessarily of Pfaff
dimension 3. Moreover, the irreversible property of turbulence decay implies
that the Euler index of the induced Cartan topology must be non-zero.

\item  The geometric concept of length scales and time scales should be
replaced by the topological concept of connectedness vs. disconnectedness.

\item  The transition to turbulence can take place only by discontinuous
solutions to the Navier-Stokes equations.

\item  The decay of turbulence can be described by continuous but
irreversible solutions to the Navier-Stokes equations. The decay of domains
for which there exists finite Topological Torsion is dependent upon a finite
viscosity, and the lack of integrability for the lines of vorticity. One
model for the decay of turbulence might be described as the collapse of
tubular domains of torsion current, becoming ever finer filamentary domains
without helicity until ultimately they have a measure zero. \ 

\item  A coherent structure in a turbulent flow may be defined as a
connected deformable component of a disconnected Cartan topology. For the
Cartan topology, these coherent structures are of two species. One component
will have null Topological Torsion and the other component will have
non-null Topological Torsion. Bounded domains for which the integral of
Topological Torsion is a flow invariant can form in a viscous turbulent
fluid. In particular, domains for which $curl\;\mathbf{v}\bullet curlcurl\;%
\mathbf{v}=0$, but $\mathbf{v}\bullet curl\;\mathbf{v}\neq 0$, can have an
evolutionary persistence in a viscous fluid. The existence of helicity
density is a sufficient but not necessary signature that the Cartan topology
is disconnected
\end{enumerate}

\section{ACKNOWLEDGMENTS}

This work was supported in part by the Energy Laboratory at the University
of Houston. The author owes the late A. Hildebrand, the director of the
Energy Lab, a debt of gratitude for first, his skepticisms, then his
conversion, and finally his insistence that the Cartan ideas of differential
forms would have practical importance in physical systems. \ 

The concepts actually began in 1975-1976 when due to the support of NASA-
Ames, the idea that the concept of Frobenius integrability delineated
streamline flows from turbulent flows was formulated [Kiehn 1976]. \ When
the Action 1-form satisfies $A\symbol{94}dA=0$, the Frobenius condition is
valid and the flow is streamline. \ When $A\symbol{94}dA\neq 0$ the
Frobenius condition for uniquew integrability fails. \ The 3-form $A\symbol{%
94}dA$ defined as topological torsion is now more popularly known in field
theories as the Chern-Simons term. \ Most of the results were presented at
an APS\ meeting in 1991.

\section{APPENDIX A : THE CARTAN TOPOLOGY}

Starting in 1899, Cartan [1899,1937,1945] developed his theory of exterior
differential systems built on the Grassmann algebraic concept of exterior
multiplication, and the novel calculus concept of exterior differentiation.
These operations are applied to sets called exterior p-forms, which are
often described as the objects that form an integrand under the integral
sign. The Cartan concepts may still seem unconventional to the engineer, and
only during the past few years have they slowly crept into the mainstream of
physics. There are several texts at an introductory level that the
uninitiated will find useful [Flanders, 1963; Bamberg, 1989; Bishop,1968;
Lovelock, 1989; Nash, 1989; Greub, 1973]. A reading of Cartan's many works
in the original French will yield a wealth of ideas that have yet to be
exploited in the physical sciences. It is not the purpose of this article to
provide such a tutorial of Cartan's methods, but suffice it to say the
''raison d'\^{e}tre'' for these, perhaps unfamiliar, but simple and useful
methods is that they permit topological properties of physical systems and
processes to be sifted out from the chaff of geometric ideas that, at
present, seem to dominate the engineering and physical sciences.

For hydrodynamics, the combination of the exterior derivative and the
interior product to form the Lie derivative acting on p-forms should be
interpreted as the fundamental way of expressing an evolution operator with
properties that are independent of geometric concepts such as metric and
connection [Kiehn, 1975a]. Cartan's Lie derivative yields the equivalent of
a convective derivative that may be used to demonstrate that the laws of
hydrodynamic evolution are topological laws, not geometric laws. This
philosophy is similar to that championed by Van Dantzig [1934] with regard
to the topological content of Maxwell's equations of electrodynamics.

The details of the Cartan Topology may be found at [Kiehn 2001], but for
purposes herein recall that there the Pfaff sequence built on a 1-form has
up to four terms defined as:\bigskip 
\begin{equation}
Topo\log ical\,\,\,ACTION:A=A_{\mu }dx^{\mu }
\end{equation}

\begin{equation*}
Topo\log ical\,\,\,\,VORTICITY:F=dA=F_{\mu \nu }dx^{\mu }\,\symbol{94}%
dx^{\nu }
\end{equation*}

\begin{equation*}
Topo\log ical\,\,\,TORSION:H=A\symbol{94}dA=H_{\mu \nu \sigma }dx^\mu \,%
\symbol{94}dx^\nu \symbol{94}dx^\sigma
\end{equation*}

\begin{equation*}
Topo\log ical\,\,\,PARITY:K=dA\symbol{94}dA=K_{\mu \nu \sigma \tau }dx^{\mu
}\,\symbol{94}dx^{\nu }\symbol{94}dx^{\sigma }\symbol{94}dx^{\tau }.
\end{equation*}
These elements of the Pfaff sequence can be used to produce a basis
collection of open sets that consists of the subsets,

\begin{equation*}
B=\{A,\,A^{c},\,H,\,H^{c}\}=\{A,\,A\cup F,\,H,\,H\cup K\}
\end{equation*}
The collection of all possible unions of these base elements, and the null
set, $\emptyset ,$ generate the Cartan topology of open sets:

\begin{equation*}
\vspace{1pt}T(open)=\{X,\,\emptyset ,\,A,\,H,A\cup F,\,H\cup K,\,A\cup
H,\,A\cup H\cup K,\,A\cup F\cup H\}.
\end{equation*}
These nine subsets form the open sets of the Cartan topology constructed
from the domains of support of the Pfaff sequence constructed from a single
1-form, $A$. \ The compliments of the open sets are the closed sets of the
Cartan topology.

\begin{equation*}
\vspace{1pt}T(closed)=\{\emptyset ,\,X,\,F\cup H\cup K,\,A\cup F\cup K,A\cup
F,\,\,H\cup K,F\cup K,F,\,A\}.
\end{equation*}

From the set of 4 ''points'' $\{A,F,H,K\}$ that make up the Pfaff sequence
it is possible to construct 16 subset collections by the process of union.
It is possible to compute the limit points for every subset relative to the
Cartan topology. The classical definition of a limit point is that a point p
is a limit point of the subset Y relative to the topology T if and only if
for every open set which contains p there exists another point of Y other
than p [Lipschutz, 1965]. The results appear in the given reference \
http://arXiv/math-ph/0101033

\vspace{1pt}\vspace{1pt}By examining the set of limit points so constructed
for every subset of the Cartan system, and presuming that the functions that
make up the forms are C2 differentiable (such that the Poincare lemma is
true, $dd\omega =0,any\,\,\omega $), it is easy to show that for all subsets
of the Cartan topology every limit set is composed of the exterior
derivative of the subset, thereby proving the claim that the exterior
derivative is a limit point operator relative to the Cartan topology. \ \
For example, the open subset, $A\cup H$, has the limit points that consist
of \ $F$ and $K.\,$

It is apparent that the Cartan topology is composed of the union of two
subsets (other than $\emptyset $ and $X$) which are both open and closed,
for $\,X=A^{c}\cup \,H^{c}$, a result that implies that the Cartan topology
is not connected, unless the Topological Torsion, $H$, and hence its
closure, vanishes. This extraordinary result has a number of physical
consequences, some of which are described in [Kiehn, 1975, 1975a].

To prove that a turbulent flow must be a consequence of a Cartan topology
that is not connected, consider the following argument: First consider a
fluid at rest and from a global set of unique, synchronous, initial
conditions generate a vector field of flow. Such flows must satisfy the
Frobenius complete integrability theorem, which requires that $A\symbol{94}%
dA=0$. The Torsion current is zero for such systems. The Cartan topology for
such systems ($H=A\symbol{94}dA=0$) is connected, and the Pfaff dimension of
the domain is 2 or less. Such domains do not support Topological Torsion.
Such globally laminar flows are to be distinguished from flows that reside
on surfaces, but do not admit a unique set of connected sychronizable
initial conditions. Next consider turbulent flows which, as the antithesis
of laminar flows, can not be integrable in the sense of Frobenius; such
turbulent domains support Topological Torsion ($H=A\symbol{94}dA\neq 0$),
and therefore a disconnected Cartan topology. The connected components of
the disconnected Cartan topology can be defined as the coherent structures
of the turbulent flow. The transition from an initial laminar state ($H=0$)
to the turbulent state ($H\neq 0$) is a transition from a connected topology
to a disconnected topology. Therefore the transition to turbulence can NOT
be continuous. However, the decay of turbulence can be described by a
continuous transformation from a disconnected topology to a connected
topology. Condensation is continuous, gasification is not.

A topological structure is defined to be enough information to decide
whether a transformation is continuous or not [Gellert, 1977]. The classical
definition of continuity depends upon the idea that every open set in the
range must have an inverse image in the domain. This means that topologies
must be defined on both the initial and final state, and that somehow an
inverse image must be defined. Note that the open sets of the final state
may be different from the open sets of the initial state, because the
topologies of the two states can be different.

There is another definition of continuity that is more useful for it depends
only on the transformation and not its inverse explicitly. A transformation
is continuous if and only if the image of the closure of every subset is
included in the closure of the image. This means that the concept of closure
and the concept of transformation must commute for continuous processes.
Suppose the forward image of a \ 1-form $A$ is $Q,$ and the forward image of
the set $F=dA$ is $Z$. Then if the closure,$\,\,A^{c}=A\cup F$ is included
in the closure of$\,\,\,\,Q^{c}=Q\cup dQ$, for all sub-sets, the
transformation is defined to be continuous. The idea of continuity becomes
equivalent to the concept that the forward image $Z$ of the limit points, $%
dA $, is an element of the closure of$\,\,Q$ [Hocking, 1961]:

\begin{quote}
\medskip

A function $f$ \ that produces an image $f[A]=Q$ is continuous iff for every
subset $A$ of the Cartan topology, $Z=f[dA]\subset Q^{c}=(Q\cup dQ)$.

\medskip
\end{quote}

The Cartan theory of exterior differential systems can now be interpreted as
a topological structure, for every subset of the topology can be tested to
see if the process of closure commutes with the process of transformation.
For the Cartan topology, this emphasis on limit points rather than on open
sets is a more convenient method for determining continuity. A simple
evolutionary process, $X\Rightarrow Y$, is defined by a map $\Phi $. The
map, $\Phi $, may be viewed as a propagator that takes the initial state, $X$%
, into the final state, $Y$. For more general physical situations the
evolutionary processes are generated by vector fields of flow, $\mathbf{V}$.
The trajectories defined by the vector fields may be viewed as propagators
that carry domains into ranges in the manner of a convective fluid flow. The
evolutionary propagator of interest to this article is the Lie derivative
with respect to a vector field, $\mathbf{V}$, acting on differential forms, $%
\Sigma \,\;$[Bishop, 1968].

\medskip

The Lie derivative has a number of interesting and useful properties.\medskip

\begin{enumerate}
\item  The Lie derivative does not depend upon a metric or a connection.

\item  The Lie derivative has a simple action on differential forms
producing a resultant form that is decomposed into a transversal and an
exact part: 
\begin{equation}
L_{(\mathbf{V})}\omega =i(V)d\omega +di(V)\omega .
\end{equation}
This formula is known as ''Cartan's magic formula''. \ \ For those vector
fields $V$ which are ''associated'' with the form $\omega ,$ such that $%
i(V)\omega =0,$ the Lie derivative becomes equivalent to the covariant
derivative of tensor analysis. \ Otherwise the two derivative concepts are
distinct.

\item  The Lie derivative may be used to describe deformations and
topological evolution. Note that the action of the Lie derivative on a
0-form (scalar function) is the same as the directional or convective
derivative of ordinary calculus, 
\begin{equation}
L_{(\mathbf{V})}\Phi =i(V)d\Phi +di(V)\Phi =i(V)d\Phi +0=\mathbf{V}\cdot
grad\Phi .
\end{equation}
It may be demonstrated that the action of the Lie derivative on a 1-form
will generate equations of motion of the hydrodynamic type.

\item  With respect to vector fields and forms constructed over C2
functions, the Lie derivative commutes with the closure operator. Hence, the
Lie derivative (restricted to C2 functions) generates transformations on
differential forms which are continuous with respect to the Cartan topology.
To prove this claim:

First construct the closure, $\{\Sigma \cup d\Sigma \}$~. Next propagate $%
\Sigma \,\ $and $d\Sigma $ by means of the Lie derivative to produce the
decremental forms, say $Q$ and $Z$, 
\begin{equation}
L_{(\mathbf{V})}\Sigma =Q\,\,\,\,\,\,\,\,and\,\,\,\,\,\,L_{(\mathbf{V}%
)}d\Sigma =Z.
\end{equation}
Now compute the contributions to the closure of the final state as given by $%
\{Q\cup dQ\}$. If $Z=dQ$, then the closure of the initial state is
propagated into the closure of the final state, and the evolutionary process
defined by $\mathbf{V}$ is continuous. \ However, 
\begin{equation*}
dQ=dL_{(\mathbf{V})}\Sigma =di(V)d\Sigma +dd(i(V)\Sigma )=di(V)d\Sigma \,\,
\end{equation*}
as$\,\,\,\,dd(i(V)\Sigma )=0\,\,\,\,$for C2 functions. But, 
\begin{equation*}
Z=L_{(\mathbf{V})}d\Sigma =d(i(V)d\Sigma )+i(V)dd\Sigma =di(V)d\Sigma
\end{equation*}
as$\,\,\,i(V)dd\Sigma =0$ for C2 p-forms. \ \ It follows that $Z=dQ$, and
therefore $\mathbf{V}$ generates a continuous evolutionary process relative
to the Cartan topology. $QED$ \ 

Certain special cases arise for those subsets of vector fields that satisfy
the equations, $d(i(\mathbf{V})\Sigma )=0$. In these cases, only the
functions that make up the p-form, ~$\Sigma $, need be C2 differentiable,
and the vector field need only be C1. Such processes will be of interest to
symplectic processes, with Bernoulli-Casimir invariants.\medskip
\end{enumerate}

By suitable choice of the 1-form of action it is possible to show that the
action of the Lie derivative on the 1-form of action can generate the Navier
Stokes partial differential equations [Kiehn, 1978]. The analysis above
indicates that C2 differentiable solutions to the Navier-Stokes equations
can not be used to describe the transition to turbulence. The C2 solutions
can, however, describe the irreversible decay of turbulence to the globally
laminar state.

For a given Lagrange Action, A, Cartan has demonstrated that the first
variation of the Action integral (with fixed endpoints) is equivalent to the
search for those vector fields, $\mathbf{V}$, that for any renormalization
factor, $\rho $, satisfy the equations

\begin{equation}
i(\rho \mathbf{V})dA=0
\end{equation}
Such vector fields are called extremal vector fields [Klein, 1962, Kiehn,
1975a]. The Cartan theorem [Cartan, 1958] states that the extremal
constraint furnishes both necessary and sufficient conditions that there
exists a Hamiltonian representation for $\mathbf{V}$, on a space of odd
Pfaff dimension (2n+1 state space). \ The resultant equations are a set of
partial differential equations that represent extremal evolution. The
renormalization condition is common place in the projective geometry of
lines, and does not require the Riemannian or euclidean concept of an inner
product or a metric [Meserve, 1983]. Hamiltonian systems are not considered
to be dissipative. \ The very strong topological constraint can be relaxed
on a space of even Pfaff dimension and still yield a Hamiltonian
representation if $\ i(\rho \mathbf{V})dA=d\theta .$ \ Moreover, it can be
shown that evolutionary processes $\mathbf{V}$ that satisfy the
Helmholtz-Symplectic constraint, (which includes all Hamiltonian processes)

\begin{equation}
d(i(\rho \mathbf{V})dA)=0,
\end{equation}
are thermodynamically reversible, relative to the Cartan topology.

A more general expression for the Cartan condition is given by the
transversal condition,

\begin{equation}
i(\rho \mathbf{V})dA=-\mathbf{f}\bullet (d\mathbf{x}-\mathbf{V}dt)-d(kT).
\label{a3}
\end{equation}
This extension of the Cartan Hamiltonian constraint is transversal for the
first term is orthogonal to the vector field, ($\mathbf{V}$). The Lagrange
multipliers, $\mathbf{f}$, are arbitrary for such a transversal constraint,
but if chosen to be of the form, $\mathbf{f}$ $=\nu \,curlcurl\,\mathbf{V}$,
then the constrained Cartan topology will generate the Navier-Stokes
equations.

As an example of this Cartan technique, substitute the 1-form of action
given by the expression,

\begin{equation}
A=\sum_{1}^{3}\mathbf{v}_{k}dx^{k}-\mathcal{H}dt,
\end{equation}
where the ''Hamiltonian'' function, H, is defined as,

\begin{equation}
\mathcal{H}=\mathbf{v\bullet v}/2+\int dP/\rho -\lambda \;div\;\mathbf{v}+kT
\end{equation}
into the constraint equation given by \ref{a3}. Carry out the indicated
operations of exterior differentiation and exterior multiplication to yield
a system of necessary partial differential equations yields of the form,

\begin{equation}
\partial \mathbf{v}/\partial t+grad(\mathbf{v\bullet v}/2)-\mathbf{v}\times
curl\mathbf{v}=-gradP/\rho +\lambda \;grad\;div\;\mathbf{v}-\nu \;curlcurl\;%
\mathbf{v.}
\end{equation}
These equations are exactly the Navier-Stokes partial differential equations
for the evolution of a compressible viscous irreversible flowing fluid.

By direct computation, the 2-form $F=dA$ has components,

\begin{eqnarray}
F &=&dA=\omega _{z}dx\symbol{94}dy+\omega _{x}dy\symbol{94}dz+\omega _{y}dz%
\symbol{94}dx \\
&&+a_{x}dx\symbol{94}dt+a_{y}dy\symbol{94}dt+a_{z}dz\symbol{94}dt,
\end{eqnarray}
where by definition

\begin{equation}
\mathbf{\omega }=curl\;\mathbf{v},\;\;\;\;\mathbf{a}=-\partial \mathbf{v}%
/\partial t-grad\mathcal{H}
\end{equation}

The 3-form of Helicity or Topological Torsion, H, is constructed from the
exterior product of A and dA as,

\begin{eqnarray}
H &=&A\symbol{94}dA=H_{ijk}dxi\symbol{94}dxj\symbol{94}dxk \\
&=&-\mathbf{T}_{x}dy\symbol{94}dz\symbol{94}dt-\mathbf{T}_{y}dz\symbol{94}dx%
\symbol{94}dt-\mathbf{T}_{z}dx\symbol{94}dy\symbol{94}dt+hdx\symbol{94}dy%
\symbol{94}dz,
\end{eqnarray}
where \textbf{T} is the fluidic Torsion axial vector current, and h is the
torsion (helicity) density:

\begin{equation}
\mathbf{T}=\mathbf{a\times v}+\mathcal{H}\mathbf{\omega },\;\;\;\;\;\;h=%
\mathbf{v\bullet \omega }
\end{equation}

The Torsion current, \textbf{T}, consists of two parts. The first term
represents the shear of translational accelerations, and the second part
represents the shear of rotational accelerations. The topological torsion
tensor, $H_{ijk}$ , is a third rank completely anti-symmetric covariant
tensor field, with four components on the variety \{x,y,z,t\}.

The Topological Parity becomes

\begin{equation}
K=dH=dA\symbol{94}dA=-2(\mathbf{a\bullet \omega })dx\symbol{94}dy\symbol{94}%
dz\symbol{94}dt.
\end{equation}
This equation is in the form of a divergence when expressed on \{x,y,z,t\},

\begin{equation}
div\mathbf{T}+\partial h/\partial t=-2(\mathbf{a\bullet \omega }),
\end{equation}
and yields the helicity-torsion current conservation law if the anomaly, $-2(%
\mathbf{a\bullet \omega })$, on the RHS vanishes. It is to be observed that
when $K=0$, the integral of $K$ vanishes, which implies that the Euler
index, $\chi $, is zero. It follows that the integral of $H$ over a boundary
of support vanishes by Stokes theorem. This idea is the generalization of
the conservation of the integral of helicity density in an Eulerian flow.
Note the result is independent from viscosity, subject to the constraint of
zero Euler index, $\chi $ = 0.

The Navier-Stokes equations of topological constraint may be used to express
the acceleration term, \textbf{a}, kinematically; i.e.,

\begin{equation}
\mathbf{a}=-\partial \mathbf{v}/\partial t-grad\mathcal{H}=-\mathbf{v}\times 
\mathbf{\omega }+\nu \;curl\;\mathbf{\omega .}
\end{equation}
Substition of this expression into the definition of the Torsion current
yields a formula in terms of the helicity density, h, the viscosity, $\nu $,
and the Lagrangian function,

\begin{equation}
\mathcal{L}=\mathbf{v\bullet v}-\mathcal{H},
\end{equation}
that may be written as:

\begin{eqnarray}
T &=&\{h\mathbf{v}-L\mathbf{\omega \}-}\nu \;\{\mathbf{v}\times curl\;%
\mathbf{\omega \}} \\
&=&\{h\mathbf{v}-(\mathbf{v\bullet v}/2-\int dP/\rho -\lambda \;div\;\mathbf{%
v)\omega \}-}\nu \;\{\mathbf{v}\times curl\;\mathbf{\omega \}}
\end{eqnarray}
Note that the torsion axial vector current persists even for Euler flows,
where $\nu $ and $\lambda $ vanish. The measurement of the components of the
Torsion current have been completely ignored by experimentalists in
hydrodynamics.

Similarly, the Topological Parity pseudo-scalar for the Navier-Stokes fluid
becomes expressible in terms of engineering quantities as,

\begin{equation}
K=2\nu (\mathbf{\omega \bullet \;}curl\mathbf{\;\omega })dx\symbol{94}dy%
\symbol{94}dz\symbol{94}dt.
\end{equation}
The Euler index for the Navier-Stokes fluid is proportional to the integral
of the Topological Parity 4-form, which is the ''top Pfaffian'' in Chern's
analysis [Chern, 1944, 1988]. When

\begin{equation}
(\mathbf{\omega \bullet \;}curl\mathbf{\;\omega })
\end{equation}
the Euler index of the induced Cartan topology must vanish. This result is
to be compared to the classic hydrodynamic principle of minimum rate of
energy dissipation [Lamb, 1932]. For a barotropic Navier-Stokes fluid of
Pfaff dimension 4, the viscosity cannot be zero, and the lines of vorticity
must be non-integrable in the sense of Frobenius.

\section{REFERENCES}

Baldwin, P. and Kiehn, R. M., 1991, ''Cartan's Topological Structure'', talk
present at the Santa Barbara summer school (1991)

Bamberg, P. and Sternberg, S., 1990, A Course in Mathematics for Students of
Physics. Vol I and II, (Cambridge University Press, Cambridge)

Berge, B., Pomeau, Y. and Vidal, C., 1984, Order within Chaos, (Wiley, NY)
p. 161.

Bishop, R. L. and Goldberg, S. I., 1968, Tensor Analysis on Manifolds,
(Dover, N. Y.) p. 199.

Bryant, R.L.,Chern, S.S., Gardner, R.B.,Goldschmidt, H.L., and Griggiths, P.
A. 1991, Exterior Differential Systems, (Springer Verlag)

Cartan, E., 1899, ''Sur certaine expressions differentielles et le systeme
de Pfaff'' Ann Ec. Norm. 16, p. 329.

Cartan, E., 1937, La Theorie des Spaces a Connexion Projective, Hermann,
Paris, (1937)

Cartan, E., 1945, ''Systems Differentials Exterieurs et leurs Applications
Geometriques'', Actualites sci. et industrielles , p .944.

Cartan, E., Lecons sur les invariants integrauxs, Hermann, Paris (1958).

Chern, S. S., 1944, ''A Simple intrinsic proof of the Gauss -Bonnet formula
for Closed Riemannian Manifolds'', Annals of Math. 45, p. 747- 752.

Chern, S.S. ,1988, ''Historical Remarks on Gauss-Bonnet'', MSRI 04808-88,
Berkeley, CA.

Croquette, V. and Williams, H., 1989, ''Nonlinear of oscillatory instability
on finite convective rolls'', in Advances in Fluid Turbulence, edited by G.
Doolen, R, Ecke, D. Holm and V. Steinberg, (North Holland, Amsterdam ) p.
300.

von Dantzig, D., 1934, ''Electromagnetism Independent of metrical
geometry'', Proc. Kon. Ned. Akad. v. Wet. 37, pp. 521-531, 644-652, 825-836 .

Flanders, H., 1963, Differential Forms, (Academic Press, New York (1963), p.
92 .

Forsyth, A. R. (1959), Theory of Differential Equations (Dover, N.Y.) p.96

Frisch, U., 1991, ''From global (Kolmogorov 1941) scaling to local
(multifractal) scaling in fully developed turbulence'' to appear in the
Proc. Roy. Soc. A

Gellert, W., 1977, et.al. Editors, The VNR Concise Encyclopedia of
Mathematics (Van Nostrand, New York ), p.686.

Greub, W., Halperin, S. and Vanstone, R., 1973, Connections, Curvature and
Cohomology, (Academic Press, New York).

Hermann, R., 1968, Differential Geometry and the Calculus of Variations,
(Academic Press, New York ), p. 254.

Hocking, J. G.,1961, Topology , (Addison Wesley, N.Y.), p.2.

Kida, S.,Yamada, M., and Ohkitani, K., 1989, '' A route to chaos and
turbulence'' in Advances in Fluid Turbulence, edited by G. Doolen, R, Ecke,
D. Holm and V. Steinberg, (North Holland, Amsterdam ) p. 116.

Kiehn, R. M., 1974, ''Modifications of Hamilton's Principle to include
Dissipation'', J. Math Phys

Kiehn, R. M., 1975, ''Submersive equivalence classes for metric fields'',
Lett. al Nuovo Cimento 14, p. 308.

Kiehn, R. M., 1975,a ''A Geometric model for supersymmetry'', Lett. al Nuovo
Cimento 12, p. 300.

Kiehn, R. M. , 1975a, ''Intrinsic hydrodynamics with applications to
space-time fluids'', Int. J. of Eng. Sci. 13, p. 941.

Kiehn, R. M., 1976, ''Retrodictive Determinism'', Int. J. of Eng. Sci. 14,
p. 749

http://www22.pair.com/csdc/pdf/retrodic.htm

Kiehn, R. M., 1990a ''Topological Torsion, Pfaff Dimension and Coherent
Structures'', in: H. K. Moffatt and T. S. Tsinober eds, Topological Fluid
Mechanics, (Cambridge University Press), 449-458 .

Kiehn, R. M., 1990b, ''Irreversible Topological Evolution in Fluid
Mechanics'' in Some Unanswered Questions in Fluid Mechanics ASME- Vol.
89-WA/FE-5, Trefethen, L. M. and Panton, R. L. Eds.

Kiehn, R.M., Kiehn, G. P., and Roberds, R. B., 1991, ''Parity and
Time-reversal symmetry Breaking, Singular Solutions and Fresnel Surfaces'',
Phys Rev A, 43 , p. 5665

Kiehn, R. M. , 1991a, ''Continuous Topological Evolution''

http://arXiv.org/abs/math-ph/0101032

Kiehn, R. M.,1991b ''Topological Parity and the turbulent state of a Navier
Stokes Fluid'', submitted to Fluid Dynamics Research

Kiehn, R. M., 1991c, ''Compact Dissipative Flow Structures with Topological
Coherence Embedded in Eulerian Environments'', in: Non-linear Dynamics of
Structures, edited by R.Z. Sagdeev, U. Frisch, F. Hussain, S. S. Moiseev and
N. S. Erokhin, (World Scientific Press, Singapore ) p.139-164.

Kiehn, R. M., 1992a ''Topological Defects, Coherent Structures and
Turbulence in Terms of Cartan's Theory of Differential Topology'' in
''Developments in Theoretical and Applied Mathematics'' Proceedings of the
SECTAM XVI conference, B. N. Antar, R. Engels, A.A. Prinaris and T. H.
Moulden, Editors, The University of Tennessee Space Institute, Tullahoma, TN
37388 USA.

Kiehn, R. M., 1992b ''Instabilty Patterns, Wakes and Topological Limit
Sets'' Poitier IUTAM conference, published in 1993 in J.P.Bonnet and M.N.
Glauser, (eds) \textit{Eddy Structure Identification in Free Turbulent Shear
Flows}, Kluwer Academic Publishers, p. 363

http://www22.pair.com/csdc/pdf/poitier.pdf

Kiehn, R. M., 1992c ''Hydrodynamic Wakes and Minimal Surfaces with Fractal
Boundaries'' Barcelona ERCOFTAC meeting in J.M. Redondo, O. Metais, (eds)
Mixing in Geophysical Flows, CIMNE, Barcelona p.52.

http://www22.pair.com/csdc/pdf/barcelon.pdf

Kiehn, R. M. 1976 NCA-2-OR-295-502

http://arXiv.org/abs/math-ph/0101033

Klein, J., 1962, Ann. Inst. AIFUB Fourier, Grenoble 12, p. 1 .

Kolmogorov, A. V. ,1941, Dolkl. Akad. Nauk SSSR, 31, p. 99 .

Lamb, H., 1937, Hydrodynamics, (Cambridge University Press) p. 617-619.

Landau, L. and Lifshitz, M, (1959) , Fluid Mechanics, (Pergamon Press,
London ) p.106.

Lipschutz, S.1965, General Topology, (Schaum, New York) p.88.

Lovelock, D. and Rund, H., 1989, Tensors, Differential Forms and Variational
Principles, (Dover, New York).

Meserve, B. E., 1983, Projective Geometry, (Dover, New York).

Nash, C. and Sen, S., 1989, Topology and Geometry for Physicists, (Academic
Press, San Diego).

Schouten, J. A. and Van der Kulk, W.,1949, Pfaff's Problem and its
Generalizations, (Oxford Clarendon Press)

Schultz, A., Kiehn, R. M., Post, E. J., and Roberds, R. B., 1979, ''Lifting
of the four fold EM degeneracy and PT asymmetry'', Phys Lett 74A, p.384.

Schuster, H. G.,1984, Deterministic Chaos, Physics Verlag (1984)

Slebodzinsky, W., 1970, Exterior Forms and their Applications, (PWN Warsaw).

Stewart, I., 1988, The Problems of Mathematics, (Oxford, NY) p 177.

Zimin, V., 1991, ''Hierarchical Models of Turbulence'' in: Non-linear
Dynamics of Structures, edited by R.Z. Sagdeev, U. Frisch, F. Hussain, S. S.
Moiseev and N. S. Erokhin, (World Scientific Press, Singapore ) p.273-288.

\end{document}